\documentclass[aps,prd,superscriptaddress,nofootinbib,amsmath,amsfonts,preprintnumbers,groupedaddress,showpacs,10pt,english]{revtex4}
\usepackage{amsmath}
\usepackage{amssymb}
\usepackage{babel}
\usepackage{wrapfig}
\usepackage{cancel}
\newcommand{\nn}{\nonumber \\}

\makeatletter

\usepackage{array,multirow,graphicx}
\usepackage{dcolumn}
\usepackage{newlfont}
\usepackage{bm}
\usepackage[colorlinks,citecolor=blue,urlcolor=blue,linkcolor=blue]{hyperref}
\usepackage[figtopcap]{subfigure}
\usepackage{color}

\begin{document}

\date{\today}

\title{New rotating  AdS/dS black holes in $\mathrm{f(R)}$ gravity}

\author{G.G.L. Nashed}
\email{nashed@bue.edu.eg}
\affiliation {Centre for Theoretical Physics, The British University in Egypt, P.O. Box
43, El Sherouk City, Cairo 11837, Egypt}

%\author{S.D. Odintsov}
%\email{odintsov@ieec.uab.es}
%\affiliation{Institut de Ci\`encies de l'Espai (ICE-CSIC/IEEC),
%Campus UAB, c. Can Magrans s/n, 08193, Barcelona, Spain}
%\affiliation{Instituci\'o Catalana de Recerca i Estudis Avan\c{c}ats (ICREA),
%Barcelona, Spain}
%\affiliation{Tomsk State Pedagogical University, 634061 Tomsk, Russia}

%\author{S. Nojiri}
%\email{nojiri@gravity.phys.nagoya-u.ac.jp}
%\affiliation{Kobayashi-Maskawa Institute for the Origin of Particles and the Universe, Nagoya University, Nagoya 464-8602, Japan}

\begin{abstract}
It is known that  general relativity (GR) theory is not consistent with the latest observations.  The modified gravity of GR  known as $\mathrm{f(R)}$  where  $\mathrm{R}$ is the Ricci scalar, is considered  to be a  good candidate  for dealing with the anomalies present in classical  GR. In this context, we study static rotating uncharged anti-de Sitter and de Sitter (AdS and dS) black holes (BHs)  using  $\mathrm{f(R)}$  theory without assuming any constraints on the Ricci scalar or on  $\mathrm{f(R)}$. We derive  BH solutions depend on  the convolution function and deviate from the AdS/dS Schwarzschild  BH solution of GR. Although the field equations have no dependence on the cosmological constant, the BHs are characterized by an effective cosmological constant that  depends on the convolution function. The asymptotic form of this BH solution depends  on the gravitational mass of the system and on extra terms that lead to BHs being different from GR BHs but to correspond to GR BHs under certain conditions.  We also investigate how these extra terms are responsible for making the singularities of the invariants milder than those of  the  GR BHs. We study some physical properties of the BHs  from the point of view of thermodynamics and show that there is an outer event horizon in addition to the inner Cauchy horizons. Among other  things  we show that our BH solutions satisfy the first law of thermodynamics. To check the stability of these BHs we use the geodesic deviations and  derive the stability conditions.  Finally,  using the odd-type mode it is shown that all the derived BHs are stable and have a radial speed equal to one.
\end{abstract}

%\pacs{04.50.Kd, 04.25.Nx, 04.40.Nr}
%\keywords{$\mathbf{F(R)}$ gravitational theory, analytic spherically symmetric BHs, thermodynamics, stability, geodesic deviation.}
\maketitle
\section{\bf Introduction}
%\end{center}
Extended gravitational theories (EGTs) have become topic of interest \cite{Capozziello2011, Capozziello:2011et,Nojiri:2010wj,Olmo:2011uz} since the discovery  that the expansion of the universe  is accelerating \cite{Riess:1998cb}. These EGTs are considered to  be a tool that can deal with   issues that GR  cannot  handle while at the same time preserving the success of GR at the solar system  as well as the astrophysical scale  \cite{Perlmutter:1998np,Riess:1998cb,Riess:2004nr,Hirata:1987hu,Dodelson:1993je,Cole:1994ab}.  There are many ways to extend GR: one is to include  a nonlinear function of the  Ricci or torsion scalars  in the Hilbert Einstein  action \cite{Schmidt:2006jt,Awad:2017ign}. The idea of a modified GR  was proposed soon  after Einstein  proposed  his GR. This was due to GR's shortcoming  in terms of renormalization, which makes it incompatible with quantum mechanics \cite{doi:10.1002/andp.19193641002,eddington_1988}. Later  Utiyama and DeWitt  showed that for GR theory to be compatible with  one--loop renormalization,    its action must contain higher--order curvature terms \cite{doi:10.1063/1.1724264,PhysRev.101.1597}. Generally, any form of the function $\mathrm{f(R)}$ must be consistent with the results of GR at the solar--system scale and  must also: (a) be free of ghosts (b) give  the correct result at the  Newtonian and post-Newtonian limit, (c) produce the correct cosmological dynamics, (d) fully resolve the Cauchy problem  and (e) produce cosmological perturbations that are  consistent with the cosmic microwave background radiation and large-scale cosmic structures \cite{Capozziello2011}.

 $\mathrm{f(R)}$ theory is considered to be an important EGTs  generic Einstein--Hilbert action by including the higher--order Ricci scalar as well as the  Ricci and  Riemann tensors or their derivatives \cite{10.1093/mnras/150.1.1,Starobinsky:1980te}.  $\mathrm{f(R)}$ has succeeded in explaining dark energy and dark matter. Much work has  therefore been done in relation to $\mathrm{f(R)}$: this has included  astrophysics and cosmological studies  \cite{Shah:2019mxn,Nojiri:2019dqc,Odintsov:2019ofr,PhysRevD.99.064049,Nascimento:2018sir,Miranda:2018jhu,Astashenok:2018bol,PhysRevD.99.063506,Elizalde:2018now,
 PhysRevD.99.064025,PhysRevD.99.104046,Bombacigno:2018tyw,Capozziello:2018ddp,Samanta:2019tjb} and   also experiments and observations that had the aim of differentiating $\mathrm{f(R)}$  from GR \cite{Starobinsky:1980te,Capozziello:2011et}. Moreover,  it is important to test $\mathrm{f(R)}$ theory in relation to black hole physics  because this theory can reproduce BHs that are different from those predicted by GR. Generally, it is more difficult  to derive BH solutions in  $\mathrm{f(R)}$  than GR because the relevant  differential equations are of the fourth order. However, by using the field equations of $\mathrm{f(R)}$ and by employing  spherically symmetric spacetime \cite{PhysRevD.74.064022,2018EPJP..133...18N,2018IJMPD..2750074N,Nashed:2018piz} under certain constraints  exact BH solutions  can be derived \cite{DeFelice:2010aj,Moon:2011hq,Larranaga:2011fv,Cembranos:2011sr,Sheykhi:2014jia,Sheykhi:2012zzc,Sawicki:2012pz,
Cognola:2005de,Sebastiani:2010kv,PhysRevD.83.029903,Hendi:2009sw,Hendi:2011eg,Hendi:2012nj,Mazharimousavi:2011nc}. The instabilities and related anti-evaporation of Schwarzschild and Reissner-Nordstr\"om  BH solutions in $\mathrm{f(R)}$ gravity has been studied \cite{Nojiri:2013su,Nojiri:2014jqa}.  Until now, no analytical  rotating  BH   or brane solutions of $\mathrm{f(R)}$ gravity in four--dimensions have  been derived. The purpose of this work was to derive new BH brane solutions and study their physical properties.

In Section \ref{S2}, a brief summary of  $\mathrm{f(R)}$  gravity theory is given. In Section \ref{S3} we derive  four-dimensional BHs with flat transverse section (Banados$\_$Teitelboim$\_$Zanelli (BTZ)-like solutions \cite{Banados:1992wn}) that depend on a convolution function. It is this function that makes  the BHs  different from those in GR: when the value of the function is zero the GR  BHs are recovered. This means that the convolution function appears  as a result of the presence of the higher--order curvature terms in $\mathrm{f(R)}$.   Although the $\mathrm{f(R)}$ field equations  do not contain a cosmological constant, we show that the  asymptotes of these BHs  behave  as anti-de Sitter or de Sitter (AdS or dS)  BHs due to the existence of an effective cosmological constant that depends on the convolution function.  The invariants of these BHs are calculated and we show that their singularities  are softer than those of GR BHs. Also in Section \ref{S3}, we apply a coordinate transformation and derive a novel rotating non-trivial  BH using  $\mathrm{f(R)}$ gravitational theory. In Section \ref{S4} we study the physical properties of this type of BH and show that the first law of thermodynamics is satisfied. In Section \ref{S5}, by using the geodesic deviation we derive the conditions for stability and illustrate  the domain of stability analytically and graphically. We reserve the final section for a discussion and the conclusion.
 \section{Fundamentals of $\mathrm{f(R)}$ gravitational theory}\label{S2}
In this section, we consider the four-dimensional action of  $\mathrm{f(R)}$ gravity, where    $\mathrm{f(R)}$ is an arbitrary differential function. It is important to stress  that   $\mathrm{f(R)}$ gravity is a modification of  GR and corresponds with the Einsteinian GR at lower order of the Ricci scalar, i.e.,   $\mathrm{f(R)=R}$. When  $\mathit{f(R)\neq R}$, we have a theory that is different from GR. The action  of $\mathrm{f(R)}$  gravity  can take the form (cf. \cite{Carroll:2003wy,1970MNRAS.150....1B,Nojiri:2003ft,Capozziello:2003gx,Capozziello:2011et,Nojiri:2010wj,Nojiri:2017ncd,Capozziello:2002rd} \ ):
\begin{eqnarray} \label{a2} {\mathop{\mathrm{ I}}}:=\frac{1}{2\kappa} \int d^4x \sqrt{-g}\,\mathrm{f(~R~)}\,,\end{eqnarray}
where $\kappa$ is  Newton's gravitational constant and  $g$ is the determinant of the metric.

Applying the variations principle  to Eq.  (\ref{a2}) gives  the vacuum field equations \cite{2005JCAP...02..010C}
 %,Koivisto:2005yc }
%\newpage
\begin{eqnarray} \label{f1}
{\mathop{\mathrm{ I}}}_{\mu \nu}=\mathrm{ R}_{\mu \nu} \mathrm{f_{R}}-\frac{1}{2}g_{\mu \nu}\mathrm{f(~R~)}+[g_{\mu \nu}\Box -\nabla_\mu \nabla_\nu]\mathrm{ f}_{_\mathrm{ R}} \equiv0,\end{eqnarray}
where  $\Box$ is the d'Alembertian operator and \[\displaystyle  \mathrm{f_{R}}=\frac{\mathrm {df}}{\mathrm {dR}}\,.\]
The trace of the field equations ~( Eq. \ref{f1}), takes the form:
\begin{eqnarray} \label{f3}
{\mathop{\mathrm{ I}}}=3\Box {\mathrm f_{R}}+\mathrm{ R}{f_{R}}-2\mathrm f(~R~)\equiv0 \,.\end{eqnarray}
From Eq. (\ref{f3} \ )  $\mathrm f(R)$  can be obtained in the form:
\begin{eqnarray} \label{f3s}
\mathrm f(~R~)=\frac{1}{2}\big[3\Box {\mathrm f_{R}}+\mathrm{ R}{f_{R}}\Big]\,.\end{eqnarray}
Substituting Eq. (\ref{f3s}) in Eq. (\ref{f1}) gives \cite{Kalita:2019xjq}
\begin{eqnarray} \label{f3ss}
{\mathop{\mathrm{ I}}}_{\mu \nu}=\mathrm{ R}_{\mu \nu} \mathrm{f_{R}}-\frac{1}{4}g_{\mu \nu}\mathrm{ R}\mathrm{ f}_{_\mathrm{ R}}+\frac{1}{4}g_{\mu \nu}\Box\mathrm{ f}_{_\mathrm{ R}} -\nabla_\mu \nabla_\nu\mathrm{ f}_{_\mathrm{ R}}=0  \,.\end{eqnarray}
Accordingly, it is  important to examine Eqs. (\ref{f3}) and (\ref{f3ss}) in the case of a flat horizon spacetime  and try to derive new BH solutions.
%%%%%%%%%%%%%%%%%%%%%%%%%%%%%%%%%%% Section 3 %%%%%%%%%%%%%%%%%%%%%%%%%%%%%%%%%%%%%%%%
\section{$AdS$ and $dS$  BH solutions for flat horizons spacetime }\label{S3}
%%%%%%%%%%%%%%%%%%%%%%%%%%%%%%%%%%%%%%%%%%%%%%%%%%%%%%%%%%%%%%%%%%%%%%%%%%%%%%%%%%%%%%
 In order to derive a general form of the arbitrary function $\mathrm{f(~R~)}$ from the equations of motion (\ref{f3}) and (\ref{f3ss}) without assuming any restrictions on the Ricci scalar, we use a flat horizon spacetime of the following form:
%%%%%%%%%%%%%%%%%%%%%%%%%%%%%%%%%%% Section 3 %%%%%%%%%%%%%%%%%%%%%%%%%%%%%%%%%%%%%%%%
%\subsection{Spherically symmetric solution}
%Assuming the spherically-symmetric the line-element to be in the form:
\begin{eqnarray} \label{met12}
& &  ds^2=-V(r)dt^2+\frac{dr^2}{W(r)}+r^2(d\phi{}^2+d\xi{}^2)\,,  \end{eqnarray}
where $-\infty\leq t \leq \infty$, \,$0\leq r \leq \infty$,\,  $0\leq \phi \leq 2\pi$,\, $-\infty\leq \xi \leq \infty$, \, and $V(r)$ and $W(r)$ are  two unknown functions that depend on  $r$.  Using Eq. (\ref{met12}) we obtain the Ricci scalar
  \begin{eqnarray} \label{Ricci}
  {\textit R(r)}=\frac{r^2WV'^2-r^2VV'W'-2r^2VWV''-4rVWV'-4V^2(W+rW')}{2r^2V^2}\,,
  \end{eqnarray}
  where $V\equiv V(r)$,  $W\equiv W(r)$,  $V'=\frac{dV}{dr}$, $V''=\frac{d^2V}{dr^2}$ and $W'=\frac{dW}{dr}$.
  Plugging Eqs.
 (\ref{f3}), (\ref{f3ss}) with Eq. (\ref{met12}) and by  using Eq. (\ref{Ricci}) we get:
 \begin{eqnarray}
&& {\mathop{\mathrm{ I}}}_t{}^t=\frac{1}{8r^2X^2}\Bigg\{r^2\Big[2X^2WF''+WFX'^2-2X^2FW''-2XWFX''\Big]-rXX'\Big\{3rWF'+F\Big[3rW'+4W\Big]\Big\}\nonumber\\
&&+2X^2\Bigl[2FW+rF'\Big\{2W-rW'\Big\}\Bigr]\Bigg\}=0\,,\nonumber\\
&&{\mathop{\mathrm{ I}}}_r{}^r=\frac{1}{8r^2X^2}\Bigg\{r^2[FWX'^2-2X^2FW''-6WX^2F''-2XWFX'']+rXX'\Big\{rWF'+F\Big[4W-3rW'\Big]\Big\}\nonumber\\
&&+2X^2\Bigl[2FW+rF'\Big\{2W-rW'\Big\}\Bigr]\Bigg\}=0\,,\nonumber\\
&&{\mathop{\mathrm{ I}}}_\phi{}^\phi={\mathop{\mathrm{ I}}}_\xi{}^\xi=\frac{1}{8r^2X^2}\Bigg\{r^2\Big[2X^2WF''+2FX^2W''+2XFWX''\Big]+rXF'\Big[rWX'-2X[2W-rW']\Big]\nonumber\\
&&-F\Big[4WX^2-3r^2XX'W'+r^2WX'^2\Big]\Big]\Bigg\}=0\,,\nonumber\\
&&{\mathop{\mathrm{ I}}}=\frac{1}{2r^2X^2}\Bigg\{r^2\Big[6X^2WF''-2X^2FW''-2FXWX''+FWX'^2\Big]+rXX'\Big[3rWF'-F\Big\{3rW'-4W\Big\}\Big]\nonumber\\
&&+6rX^2F'\Big[2W+rW'\Big]+2rFW'+FW+r^2f(r)\Bigg\}=0\,,
\label{feq}
\end{eqnarray}
where $X(r)=\frac{V(r)}{W(r)}$ and $F\equiv F(r)=\frac{\mathrm{df(R(r))}}{\mathrm{dR(r)}}$, $F'=\frac{\mathrm{dF(r)}}{\mathrm{dr}}$, $F''=\frac{\mathrm{d^2F(r)}}{\mathrm{dr^2}}$ and $F'''=\frac{\mathrm{d^3F(r)}}{\mathrm{dr^3}}$. Since we are dealing with flat horizon spacetime in which the metric potentials depend on the radial coordinate, we take $\mathrm{f(R)}=\mathrm{f(r)}$.

Omitting the part that describes the trace Eq.  (\ref{feq}) can be rewritten in the following form:
\begin{align}
\label{E1n}
0=& r^2\Big[2X^2WF''+WFX'^2-2X^2FW''-2XWFX''\Big]-rXX'\Big\{3rWF'+F\Big[3rW'+4W\Big]\Big\}\nonumber\\
&+2X^2\Bigl[2FW+rF'\Big\{2W-rW'\Big\}\Bigr]\, , \\
\label{E2n}
0=&r^2[FWX'^2-2X^2FW''-6WX^2F''-2XWFX'']+rXX'\Big\{rWF'+F\Big[4W-3rW'\Big]\Big\}\nonumber\\
&+2X^2\Bigl[2FW+rF'\Big\{2W-rW'\Big\}\Bigr]\, , \\
\label{E3n}
0=&r^2\Big[2X^2WF''+2FX^2W''+2XFWX''\Big]+rXF'\Big[rWX'-2X[2W-rW']\Big]\nonumber\\
&-F\Big[4WX^2-3r^2XX'W'+r^2WX'^2\Big]\Big] \, .
\end{align}
Substituting Eq.  (\ref{E2n}) from  (\ref{E1n}) then gives
\begin{equation}
\label{E4n}
0 = 2rXF''-X'(2F+rF') \, .
\end{equation}
We can also obtain the same equation  (\ref{E4n}) by adding Eq. (\ref{E1n}) to (\ref{E3n}). This clearly shows   that
only two out of these three equations  Eqs.  (\ref{E1n}), (\ref{E2n}), and (\ref{E3n}) are independent. For example Eq.~(\ref{E1n}) is equal to minus Eq.~(\ref{E2n}) minus two multiplied by  Eq.~(\ref{E3n}). This means that  we can choose for example Eq.~(\ref{E1n}) and Eq.~(\ref{E4n}) as the independent equations. Because we have three unknown functions $V$, $W$ and $F$, will not be possible to determine one  of these functions.

As an example, we derive the AdS/dS Schwarzschild--type solution by assuming that,
\begin{equation}
\label{ES1}
X=1 \,,
\end{equation}
which gives
\begin{equation}
\label{ES2n}
F''=0 \, , \quad \mbox{that is,} \quad
F=f_0 + f_1 r \, .
\end{equation}
substituting Eqs. (\ref{ES1}) and (\ref{ES2n}) into Eq.~(\ref{E1n}) we obtain
\begin{align}
\label{ES3n}
0=& 2W[f_0+2rf_1]- r^2\left[ f_1W'+W''(f_0+rf_1) \right]  \, .
\end{align}
The above equation has the following solution when  $f_1=0$:
\begin{equation}
\label{ES4n}
W(r)=w_0r^2+\frac{w_1}{r}  \, ,
\end{equation}
where $w_0$ and $w_1$ are constants.
Eq. (\ref{ES4n}) expresses the Schwarzschild--AdS/dS spacetime.

In the case where $f_0=0$  Eq.~(\ref{ES3n})  gives the following solution:
\begin{equation}
\label{ES7n}
W= {\tilde w}_0 r^2 + \frac{{\tilde w}_1 }{r^2} \, ,
\end{equation}
where ${\tilde w}_0$ and ${\tilde w}_1$ are constants. The solutions given by Eqs. (\ref{ES4n}) and (\ref{ES7n}) give a Ricci scalar that has a constant value.

When  either $f_0$ or $f_1$ does not vanish, and  when $r$ is small, the $f_0$ term in Eq. (\ref{ES3n})
dominates and the solution should  behave as Eq. (\ref{ES4n}),  when $r$ is large, the $f_1$ term in Eq. (\ref{ES3n}) dominates and the solution should behave  as Eq. (\ref{ES7n}).
This means that Eq. (\ref{ES4n}) gives a black hole solution for the region  where $r$ is  small  and Eq. (\ref{ES7n}) gives solution where  $r$ is large.

We can also consider a more general case.
By again assuming $W\neq 0$,   Eq.~(\ref{E4n}) can be rewritten as
\begin{equation}
\label{E7n}
X = \exp \left( \int \frac{2r F''}{rF' + 2 F} dr \right) \, .
\end{equation}
By substituting Eq.~(\ref{E7n}) into  Eq.~(\ref{E1n}), we then obtain:
\begin{align}
\label{E9n}
0=& -\frac{ \left(4rFF'+r^2F'^2+4F^2\right) FW''}{4(2F+rF')^2} \nn
&-\frac{ \left(3r^2FF'F''+6rF^2F''+4F^2F'+4rFF'^2+r^2F'^3\right) W'}{4(2F+rF')^2}\nn
&-\frac{ \left(2r^3FF'F''+2r^4FF'F'''+8r^2F^2F''+4r^3F^2F'''-16rF^2F'-10r^2FF'^2+2r^4F'^2F''-2r^3F'^3-8F^3\right) W}{4r^2(2F+rF')^2} \, .
\end{align}
Equation (\ref{E9n}) is a  linear homogeneous differential equation for $W$.
For example, if we take  $F\propto r^n$ where $n$ is a constant then Eq.~(\ref{E9n}) reduces to
\begin{equation}
\label{E99n}
0=   W''+ \frac{n(4 n -1) }{r(n + 2)} W'
+ \frac{4 (n^2+n+1)(n^2-2n-2) }{r^2\left( n + 2 \right)^2}W \, .
\end{equation}
The solution to Eq.~(\ref{E99n}) is given by
\begin{equation}
\label{E11}
W = w_+ r^{\alpha_1} + w_- r^{\alpha_2} \,,
\end{equation}
where $\alpha_1$ and $\alpha_2$  are constants given by
\begin{equation}
\label{E22}
\alpha_1 = \frac{-2(n^2+n + 1)}{ (n+2)}\, \qquad \qquad {\textrm and } \qquad \qquad \alpha_2 = \frac{-2(n^2-2n -2)}{ (n+2)}\, .
\end{equation}
The solution $n=0$ gives $\alpha_{1,2} =-1$, $2$ and corresponds to
the Schwarzschild--AdS/dS spacetime
given by Eq. (\ref{ES4n}) however, other cases correspond to new kinds of flat horizon spacetime solutions.

\subsection{A New BH types}
As  discussed above, we have two independent differential equations with three unknowns. Therefore, to solve these differential equations, we assume the  unknown function,  $F$ , to be
\begin{eqnarray} \label{ass1n}
F=1+\frac{a}{r^2}\,.
\end{eqnarray}
Equation (\ref{ass1n}) shows that when $a=0$ we get the GR limit since $\mathrm{f(R)}=\mathrm{cons}$. substituting Eq.~(\ref{ass1n}) in Eq.~ (\ref{feq}) we then obtain
\begin{eqnarray} \label{ass1}
&&W(r)=\frac{e^{\frac{3a}{2r^2}}}{r}\left\{a_2\mathbb{H}+a_3\mathbb{H}_1\right\}\,,\qquad X(r)=a_1e^{\frac{-3a}{r^2}}\,,\qquad  \qquad V(r)=X(r)W(r)\,,\qquad  \qquad F=1+\frac{a}{r^2}\,,\nonumber\\
&&
\end{eqnarray}
where $\mathbb{H}=\mathrm{HeunC}(\ \frac{3}{2},\frac{3}{2},0,\frac{3}{8},\frac{9}{8},-\frac{a}{r^2} \ )$ and   $\mathbb{H}_1=\mathrm{HeunC}(\ \frac{3}{2},-\frac{3}{2},0,\frac{3}{8},\frac{9}{8},-\frac{a}{r^2} \ )$\footnote{The $\mathrm{HeunC}$ function is the solution of the Heun Confluent equation which is defined as
\begin{eqnarray} \label{sp1}
Y''(r)-\frac{1+\beta-(\alpha-\beta-\gamma-2)r-r^2\alpha}{r(r-1)}Y'(r)-\frac{\alpha(1+\beta)-\gamma-2\eta-(1+\gamma)\beta-r(2\delta+[2+\gamma+\beta])}
{2r(r-1)}Y(r)=0\,.
\end{eqnarray}
The solution of the above differential equation is defined $\mathrm{HeunC}(\alpha,\beta,\gamma,\delta,\eta,r)$. For more details, interested readers can check  \cite{RONVEAUX2003177,MAIER2005171}.\\ $\mathrm{HeunCPrime}$ is the derivative of the Confluent Heun  function.}.
Substituting Eq. (\ref{ass1}) into the trace equation, i.e., the fourth equation of Eq. (\ref{feq}\ ), we obtain $\mathrm{f(r)}$ in  the form
  \begin{eqnarray} \label{ass11}
&& \mathrm{f(r)}=-\frac{2e^{^{\frac{3a}{2r^2}}}}{r^7}\Bigg\{2a(r^2+3a)[a_3r^3\mathbb{H}_1+a_2\mathbb{H}]+3r^2[a_3r^3(r^2+a)\mathbb{H}_1-2aa_2\mathbb{H}] \Bigg\},
 \end{eqnarray}
 where $a$, $a_1$, $a_2$ and $a_3$ are constants.
Substituting Eq. (\ref{ass11}) into  Eq. (\ref{Ricci}) we then get
\begin{eqnarray} \label{sol11}
&&\mathrm{R}=-\frac{2e^{^{\frac{3a}{2r^2}}}}{r^5(a+r^2)}\Bigg\{3a_3r^5(2r^2+3a)\mathbb{H}_1+a\Big[(3a+2r^2)\Big\{a_2\mathbb{H}+a_3r^3\mathbb{H}_1\Big\}-a_2r^2\mathbb{H}\Big]
\Bigg\}\,.
\end{eqnarray}
Equations (\ref{ass1}), (\ref{ass11}) and (\ref{sol11}) show that when $a=0$ we have
\begin{eqnarray}\label{reda}
 X(r)=a_1, \qquad \qquad V(r)=W(r) \qquad \qquad \mathrm{ and} \qquad \qquad F(r)=1.\end{eqnarray} Equation (\ref{reda}) shows that, when $\mathrm{F(\ r \ )}=1$, $\mathrm{f(R)}=\mathrm{R}$ also in this case, $V(r)=W(r)=a_3r^2+\frac{a_2}{r}$. All of  the above data ensure that when $a=0$ we recover to the GR--BHs\footnote{Note that, when $a=0$, we get $\mathbb{H}=\mathbb{H}_1=\mathrm{HeunC}(\frac{3}{2},\frac{3}{2},0,\frac{3}{8},\frac{9}{8},0)=\mathrm{HeunC}(\frac{3}{2},
 -\frac{3}{2},0,\frac{3}{8},\frac{9}{8},0)=1$\cite{RONVEAUX2003177,MAIER2005171}.}.

 Finally,  we note that this is the first time that the BHs described by Eq. (\ref{ass1}) have been derived and  reduce to the GR BHs when the constant $a$ vanishes.
%%%%%%%%%%%%%%%%%%%%%%%%%%%%%%%%%%% Section 3 %%%%%%%%%%%%%%%%%%%%%%%%%%%%%%%%%%%%%%%%
\subsection{Physical properties of the BHs given by equation (\ref{ass1})}
%%%%%%%%%%%%%%%%%%%%%%%%%%%%%%%%%%%%%%%%%%%%%%%%%%%%%%%%%%%%%%%%%%%%%%%%%%%%%%%%%%%%%%
In this section,  the physical properties of the BHs described by Eq. (\ref{ass1}) will be investigated. With this in mind,  we  require the asymptotic  behavior of the metric potentials $V(r)$ and $W(r)$  in (\ref{ass1}):
 \begin{eqnarray}\label{mpab}
&& V(r)\approx \mp\Lambda_{\mathrm eff} r^2-\frac{2M}{r}+\frac{3aM}{r^3}+\frac{87a^2M}{28r^5}+\cdots\,,\nonumber\\
 &&
W(r)\approx  \mp\Lambda_{\mathrm eff} (r^2+3a)-\frac{2M}{r}\pm\frac{9a^2\Lambda_{\mathrm eff}}{2r^2}-\frac{3aM}{r^3}\mp\frac{9a^3\Lambda_{\mathrm eff}}{2r^4}+\cdots\,,\end{eqnarray}
where we have assumed $a_1=1$, $a_3=\Lambda_{\mathrm eff}$ and $a_2=-2M$. Substituting  Eq. (\ref{mpab}) into Eq.  (\ref{met12}) we get
\begin{eqnarray} \label{metaf}
& &  ds^2=-\Bigg[\mp\Lambda_{\mathrm eff} r^2-\frac{2M}{r}+\frac{3aM}{r^3}+\frac{87a^2M}{28r^5}\Bigg]dt^2+\frac{dr^2}{ \mp\Lambda_{\mathrm eff} (r^2+3a)-\frac{2M}{r}\pm\frac{9a^2\Lambda_{\mathrm eff}}{2r^2}-\frac{3aM}{r^3}\mp\frac{9a^3\Lambda_{\mathrm eff}}{2r^4}}+r^2(d\phi^2+d\xi^2)\,, \nonumber\\
 && \end{eqnarray}
which   asymptotically approaches AdS/dS spacetime and does not correspond to the Schwarzschild--AdS/dS BH of GR  because of the contribution of the extra terms that come mainly from the constant parameter $a$ whose source is the effect of the higher-order curvature terms of  $\mathrm{f(R)}$.  It  can easily be checked that, when these extra terms equal zero the situation smoothly returns to the Schwarzschild spacetime \cite{Misner:1974qy}.

Next we substitute Eq. (\ref{mpab})  into Eq. (\ref{Ricci})  and   obtain
\begin{eqnarray} \label{R1}
&&R(r)\approx \pm12\Lambda_{\mathrm eff}\pm\frac{30a\Lambda_{\mathrm eff}}{r^2}\pm\frac{24a^2\Lambda_{\mathrm eff}}{r^3}+\cdots\,,\nonumber\\
 &&r(R)=\mp\frac{10a\Lambda_{\mathrm eff}}{\mathbb{R}}-\frac{\mathbb{R}}{\mathbb{R}\mp12\Lambda_{\mathrm eff}}\approx a_4+a_5\mathrm{R}+a_6\mathrm{R^2}+a_7\mathrm{R^3}+\cdots \,,
\end{eqnarray}
where $\mathbb{R}=\mp2a\Lambda_{\mathrm eff}\Big(6a-\sqrt{2a(\mp125\Lambda_{\mathrm eff}+18a R\mp216a\Lambda_{\mathrm eff})}(R\mp12\Lambda_{\mathrm eff})^{3/2}\Big)^{1/3}$ and $a_4$ $\cdots$ $a_7$ are constants.
Equation (\ref{R1}) shows that, when the constant $a\neq 0$, we have a non-trivial value of the Ricci scalar as a result the contributions from  higher--order curvature, also when $a=0$ we get a trivial value of the Ricci scalar that corresponds to GR BHs. The asymptotic form of  $\mathrm{f(r)}$,  which is given by Eq. (\ref{ass11}), has the form
\begin{eqnarray} \label{fR1}
&&f(r)\approx \pm6\Lambda_\mathrm{eff}+\pm\frac{24a\Lambda_\mathrm{eff}}{r^2}\pm\frac{12a^2\Lambda_\mathrm{eff}}{r^3}\pm\frac{45a^2\Lambda_\mathrm{eff}}{r^4}+\cdots\,.
\end{eqnarray}
Substituting the second equation in Eq. (\ref{R1}) into (\ref{fR1}) we get
\begin{eqnarray} \label{fR2}
&&f(R)\approx a_8+a_9R+a_{10}R^2+a_{11}R^3+\cdots\,,\nonumber\\
 \end{eqnarray}
 where  $a_i, i=8\cdots 11$ are constants.

Equation  (\ref{ass1}) is now used  to calculate the invariants  and we obtain
 \begin{eqnarray} \label{inv}
&& \mathrm{R}_{\mu \nu \rho \sigma} \mathrm{R}^{\mu \nu \rho \sigma}= \mp24\Lambda_\mathrm{eff}{}^2+\frac{72a\Lambda_\mathrm{eff}{}^2}{r^2}+\frac{108a^2\Lambda_\mathrm{eff}{}^2}{r^4}
\mp\frac{96a\Lambda_\mathrm{eff}}{r^5}+\frac{12(4M^2+27a^3\Lambda_\mathrm{eff}{}^2)}{r^6}\cdots\,, \nonumber\\
 &&
 \mathrm{ R}_{\mu \nu } \mathrm{R}^{\mu \nu }=36\Lambda_\mathrm{eff}{}^2+\frac{108a\Lambda_\mathrm{eff}{}^2}{r^2}+\frac{108a^2\Lambda_\mathrm{eff}{}^2}{r^4}\mp
  \frac{144aM\Lambda_\mathrm{eff}}{r^5}+\frac{162a^3\Lambda_\mathrm{eff}{}^2}{r^6}+\cdots \nonumber\\
 &&\mathrm{R}=\pm12\Lambda_{\mathrm eff}\pm\frac{30a\Lambda_{\mathrm eff}}{r^2}\pm\frac{24a^2\Lambda_{\mathrm eff}}{r^3}+\cdots,
 \end{eqnarray}
  where  $\Big(\mathrm{R}_{\mu \nu \rho \sigma} \mathrm{R}^{\mu \nu \rho \sigma}, \mathrm{R}_{\mu \nu} \mathrm{R}^{\mu \nu}, \mathrm{R}\Big)$ are the Kretschmann
scalar, the Ricci tensor square and the Ricci scalar, respectively, and all of these have a true singularity at $r=0$. It is important to stress that  the constant $a$ is the main source for the deviation of the above results from GR. The invariants of GR have the following values $\Big(\mathrm{R}_{\mu \nu \rho \sigma} \mathrm{R}^{\mu \nu \rho \sigma}, \mathrm{R}_{\mu \nu} \mathrm{R}^{\mu \nu}, \mathrm{R}\Big)=(48M^2/r^6,36\Lambda^2,\pm12\Lambda)$.
 Equation (\ref{inv}) indicates that the leading term among the invariants  $(\mathrm{R}_{\mu \nu \rho \sigma} \mathrm{R}^{\mu \nu \rho \sigma},\mathrm{R}_{\mu \nu} \mathrm{R}^{\mu \nu},\mathrm{R})$ is $(\frac{1}{r^2},\frac{1}{r^2},\frac{1}{r^2})$: this is different from  the case of the  Schwarzschild--AdS/dS BHs where the leading term of the Kretschmann
scalar as  $\frac{1}{r^\mathrm{6}}$ and the other invariants  $\mathrm{R}_{\mu \nu} \mathrm{R}^{\mu \nu}=\mathrm{R}$=constant. Therefore, Eq. (\ref{inv}) indicates that the  Kretschmann   singularity  is milder  than that of the Schwarzschild--AdS/dS BHs of GR.
%%%%%%%%%%%%%%%%%%%%%%%%%%%%%%%%%%% Section 3 %%%%%%%%%%%%%%%%%%%%%%%%%%%%%%%%%%%%%%%%
\subsection{A\lowercase{d}S/\lowercase{d}S  rotating BHs with flat horizons}\label{S31}
%%%%%%%%%%%%%%%%%%%%%%%%%%%%%%%%%%%%%%%%%%%%%%%%%%%%%%%%%%%%%%%%%%%%%%%%%%%%%%%%%%%%%%
To include the angular momentum of the BHs described by Eq. (\ref{ass1}) we apply the following transformations\footnote{It is  known that  the addition
of cosmological constant leads to the  reproduction of different kinds of rotating
BHs  \cite{Klemm:1997ea,Lemos:1994xp,Awad:2002cz}}
\begin{equation}\label{t1}
\bar{\phi} =-\Omega~ {\phi}+\frac{ \omega}{l^2}~t,\qquad \qquad \qquad
\bar{t}= \Omega~ t-\omega~ \phi\,,
\end{equation}
where $l=\frac{\mp3}{\Lambda_\mathrm{eff}}$.
Applying the above transformation  to the spacetime described by (\ref{met12}) we get
\begin{equation}\label{m1}
ds^2=-V(r)\left[\Omega d\bar{t}  -  \omega d\bar{\phi}
\right]^2+\frac{dr^2}{W(r)}+\frac{r^2}{l^4} \left[\omega
d\bar{t}-\Omega l^2 d\bar{\phi} \right]^2+ r^2d\xi^2,
\end{equation}
where $V(r)$ and $W(r)$ are given by Eq. (\ref{ass1}) and $\Omega$ is defined as
\[\Omega:=\sqrt{1\mp\frac{\omega^2}{l^2}}\,.\]
%%%%%%%%%%%%%%%%%%%%%%%%%%%%%%%%%% Section 4 %%%%%%%%%%%%%%%%%%%%%%%%%%%%%%%%%%%%%%%%
\section{Thermodynamics of the BH\lowercase{s}  }\label{S4}
\begin{figure}
\centering
\subfigure[~The metric potential of BH (\ref{mpab})]{\label{fig:met}\includegraphics[scale=0.3]{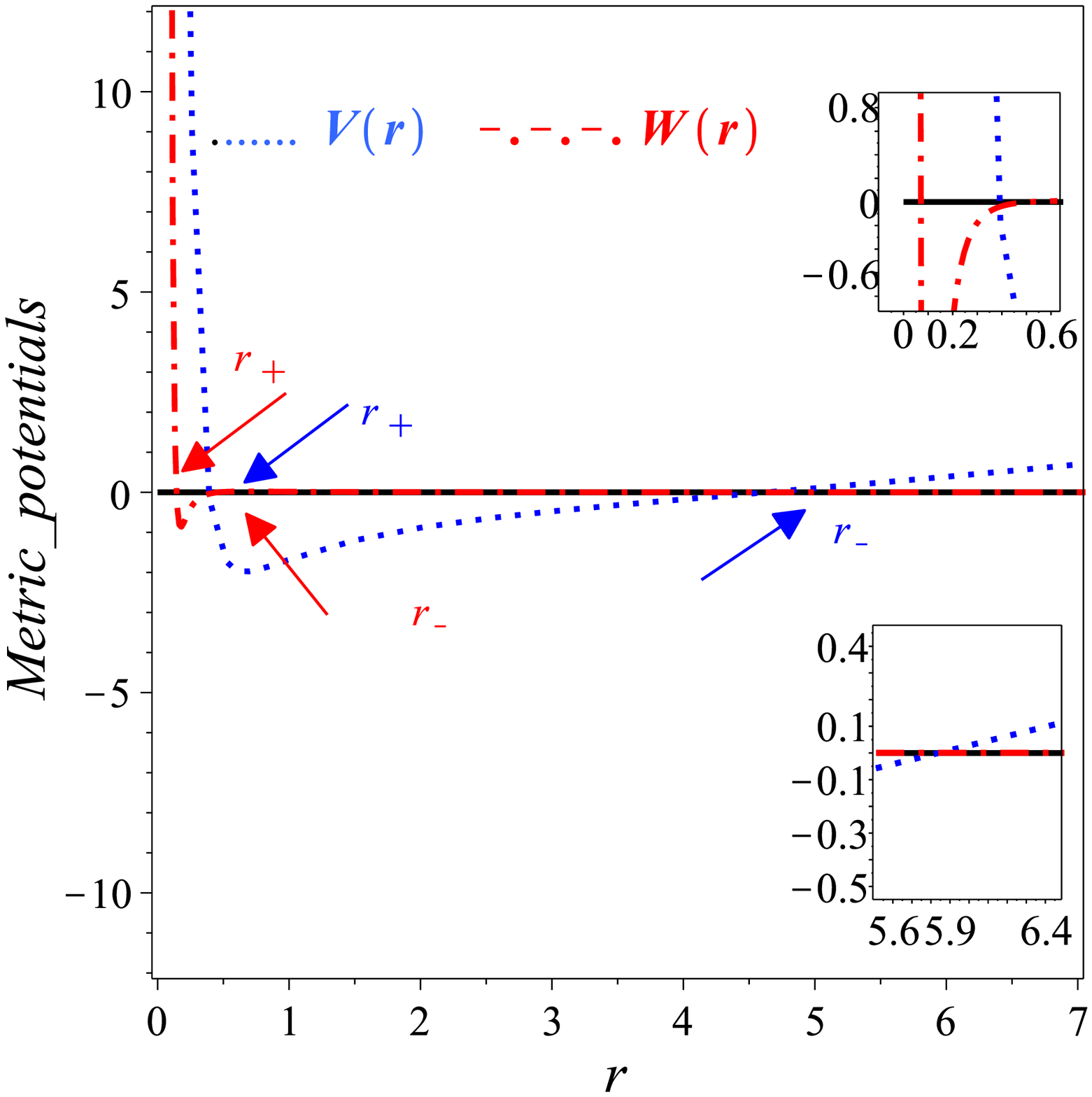}}
\subfigure[~The metric potential of BH (\ref{mpab})]{\label{fig:metrd}\includegraphics[scale=0.3]{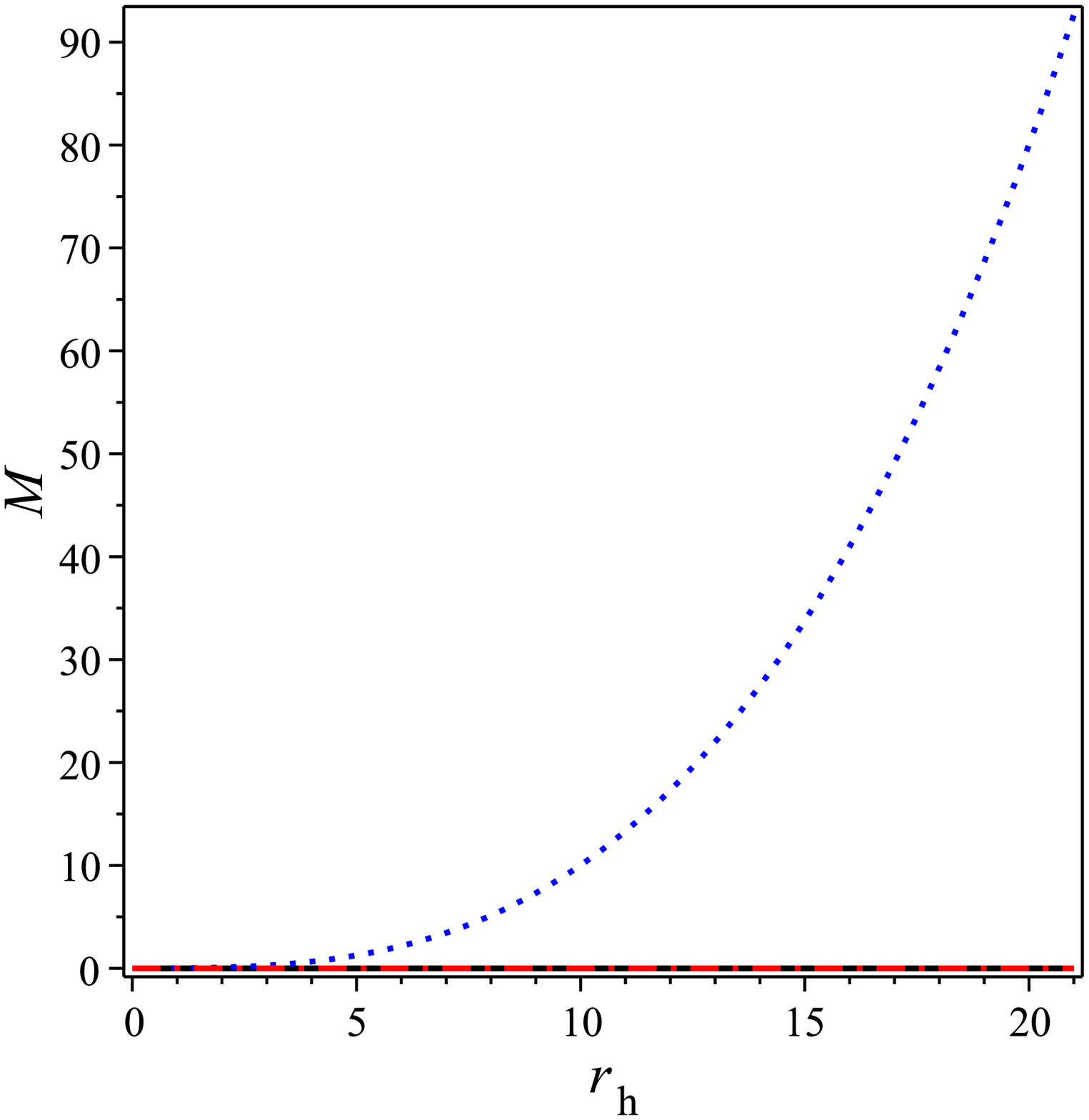}}
\subfigure[~Hawking temperature of BH (\ref{mpab})]{\label{fig:Temp}\includegraphics[scale=0.3]{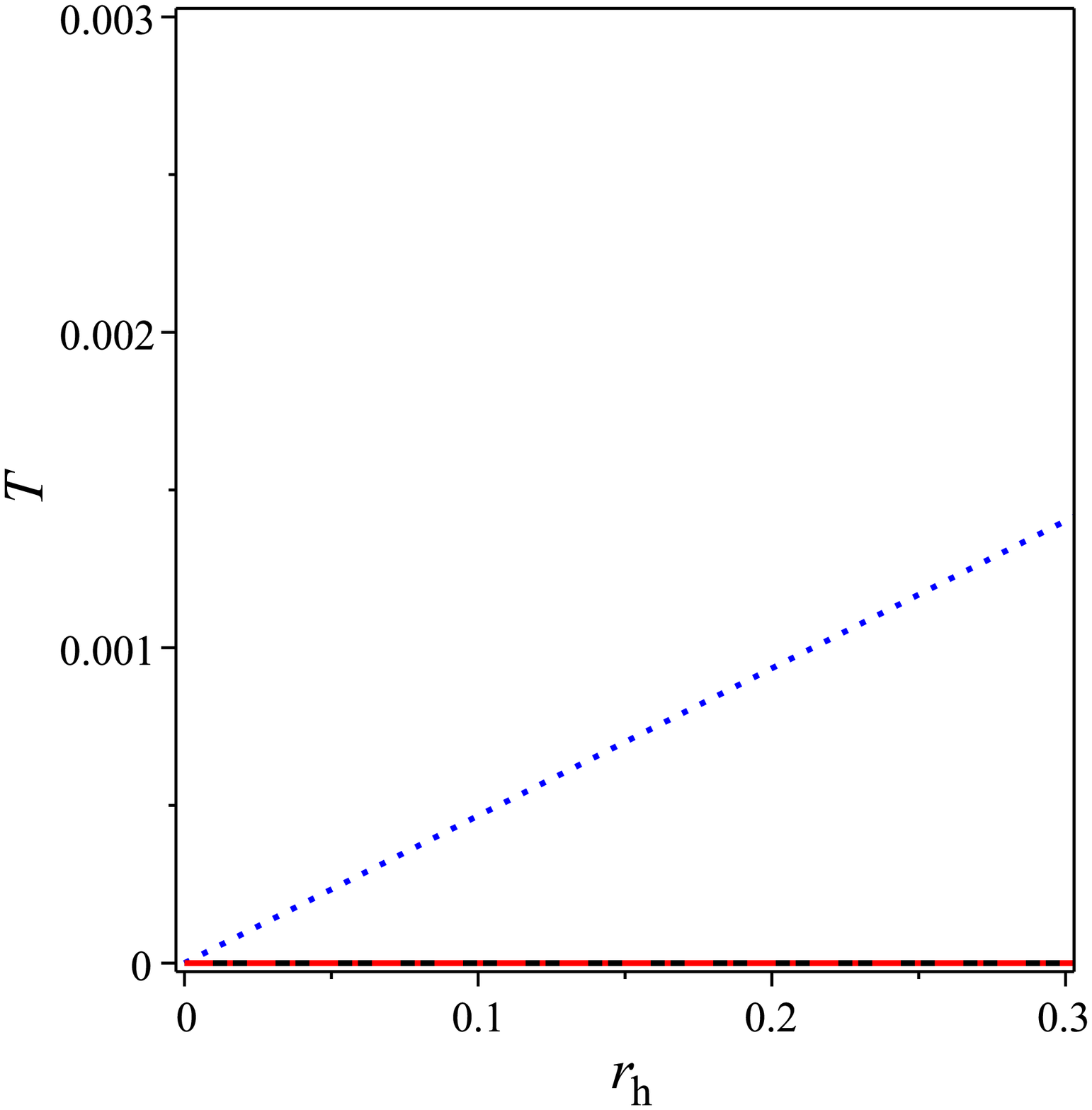}}
\caption[figtopcap]{\small{{Schematic plots of thermodynamic quantities of the black hole solution   given by Eq. (\ref{metaf}): \subref{fig:met}  the behavior of the metric potentials $V$ and $W$; \subref{fig:metrd} typical behavior of the horizons of the metric potential $V(r)$ given by Eq. (\ref{r1}); \subref{fig:Temp} typical behavior of the horizon temperature given by Eq.  (\ref{T1}), showing how the temperature increases with increasing $r_h$. All of these values are for $\Lambda_\mathrm{eff}=0.02.$}}}
\label{Fig:1}
\end{figure}

%%%%%%%%%%%%%%%%%%%%%%%%%%%%%%%%%%%%%%%%%%%%%%%%%%%%%%%%%%%%%%%%%%%%%%%%%%%%%%%%%%%%%%
The basic definitions used in  thermodynamics can be used to study the physical properties  of the BHs described by Eq. (\ref{mpab}). The Hawking  temperature is defined as \cite{PhysRevD.86.024013,Sheykhi:2010zz,Hendi:2010gq,PhysRevD.81.084040,Wang:2018xhw,Zakria:2018gsf}
\begin{equation}\label{temp}
T = \frac{V'}{4\pi}\,,
\end{equation}
where $V'$ is the derivative of the metric potential $V$ w.r.t.  the radial coordinate $r$.
The Hawking entropy of the horizons is given by
\begin{equation}\label{ent}
S=\frac{{\cal A} F}{4}\,,
\end{equation}
where ${\cal A}$  is the area of the horizon and $F$ is the derivative of $\mathrm{f(R)}$ w.r.t. $\mathrm{R}$.  The  quasi--local--energy  is given by \cite{PhysRevD.84.023515,PhysRevD.86.024013,Sheykhi:2010zz,Hendi:2010gq,PhysRevD.81.084040,Zheng:2018fyn}
\begin{equation}\label{en}
E(r)=\frac{1}{4}\displaystyle{\int }\Bigg[2\mathrm{f_{R}}(r)+r^2\Big\{\mathrm{f(R(r))}-\mathrm{R(r)}\mathrm{f_{R}(r)}\Big\}\Bigg]dr\equiv\frac{1}{4}\displaystyle{\int }\Bigg[2F+r^2\Big\{\mathrm{f(R(r))}-\mathrm{R(r)}F\Big\}\Bigg]dr\,,
\end{equation}
and the heat capacity is defined as
\begin{equation} \label{hc}
C(r)=\frac{\partial M}{\partial T}=\frac{\partial M}{\partial r}\frac{\partial r}{\partial T}\,.
\end{equation}
Finally,   the Gibbs free energy  is defined as \cite{Zheng:2018fyn,Kim:2012cma}
\begin{equation} \label{enr}
G(r)=E(r)-T(r)S(r).%+P(r_+)V(r_+),
\end{equation}
A BH described by Eq. (\ref{mpab}) can be  described using the effective cosmological constant $\Lambda_\mathrm{eff}$, the mass of the BH $M$,  and the parameter $a$. When  $a$  vanishes we obtain the Schwarzschild--AdS/dS spacetime, which corresponds to GR. To derive the horizons of the BH  (\ref{mpab}) we set $V(r)=0$ and neglect the term $O\Big(\frac{1}{r^5}\Big)$ and higher--orders to get
\begin{eqnarray}\label{r1}
&&M=\frac{\Lambda_\mathrm{eff}\, r_h{}^5}{3a-2r_h{}^2}\,.
\end{eqnarray}
The metric potentials of the type BH described by Eq. (\ref{mpab}) are illustrated   in Fig. \ref{Fig:1} \subref{fig:met}. From Fig. \ref{Fig:1} \subref{fig:met}, the  two horizons of the metric potentials $V(r)$ and  $W(r)$ can easily be seen. {In the frame of GR and modified gravitational theories,  several explicit examples of the actions which give solutions describing the non-singular BH spacetime with multi-horizons derived when coupling with non-linear electromagnetism is presented \cite{Nojiri:2017kex}.  Here in this study, we considered the linear form of the Maxwell field equation and show that the resulting  BH has two horizons only.  However, if we consider the non-linear form of the Maxwell field equation, maybe we can get a BH solution having multiple horizons. This will be addressed  elsewhere. }

 In addition the behavior of the  horizon given by Eq. (\ref{r1}) is illustrated in Fig. \ref{Fig:1} \subref{fig:metrd}.
 Eq. (\ref{temp}), the Hawking temperature  can be calculated as
\begin{eqnarray}\label{T1}
T=\frac{\Lambda_\mathrm{eff} r_h(5a-2r_h)}{4\pi(3a-2r_h)}\,.
\end{eqnarray}
The behavior of the Hawking temperature  given by Eq. (\ref{T1}) is illustrated in Fig. \ref{Fig:1} \subref{fig:Temp} which shows that the Hawking temperature is always positive and  increasing as $r_h$ increases.

\begin{figure}
\centering
\subfigure[~Entropy of BH (\ref{mpab})]{\label{fig:ent}\includegraphics[scale=0.3]{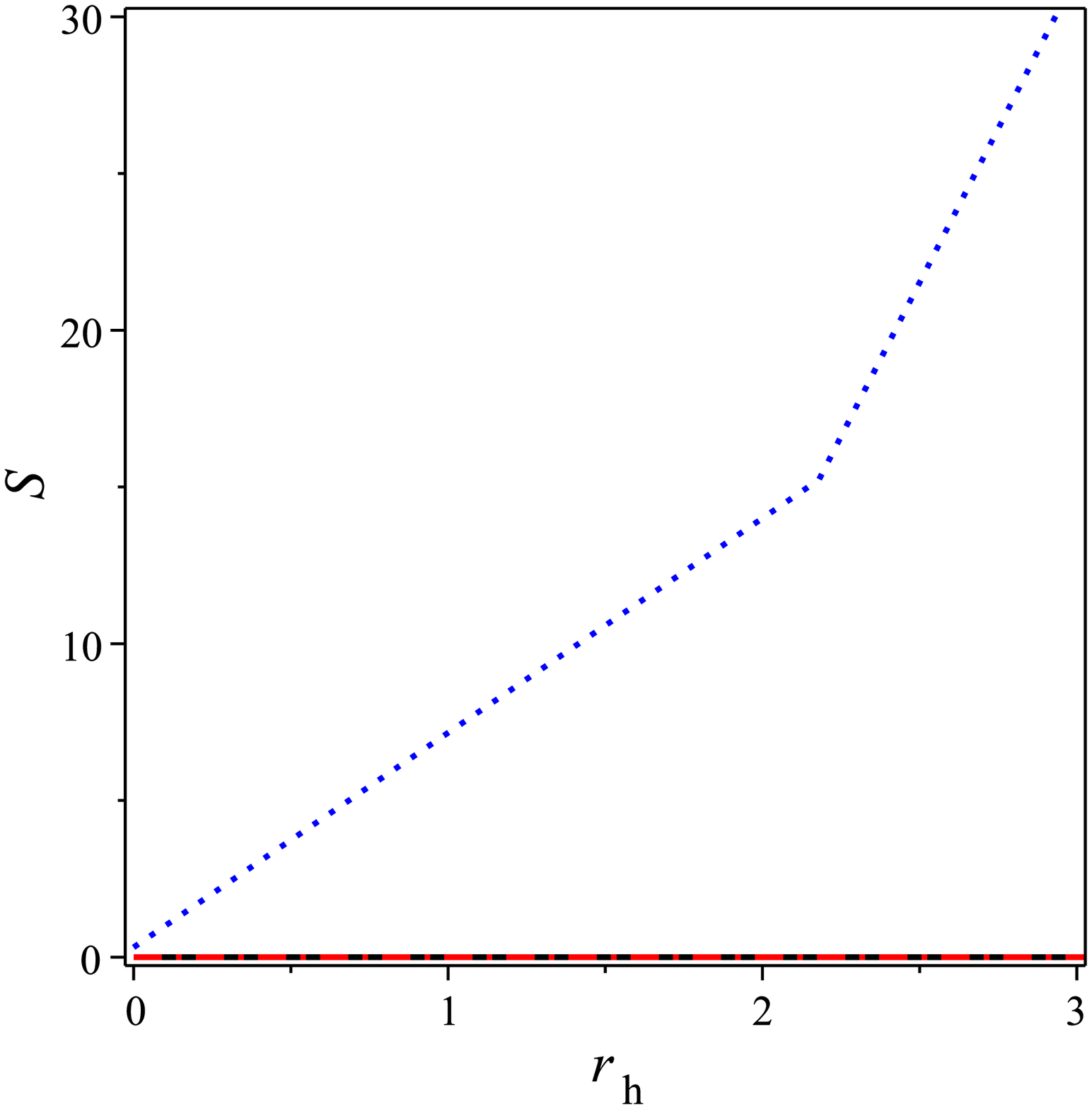}}
\subfigure[~Quasi local energy of BH (\ref{mpab})]{\label{fig:Enr}\includegraphics[scale=0.3]{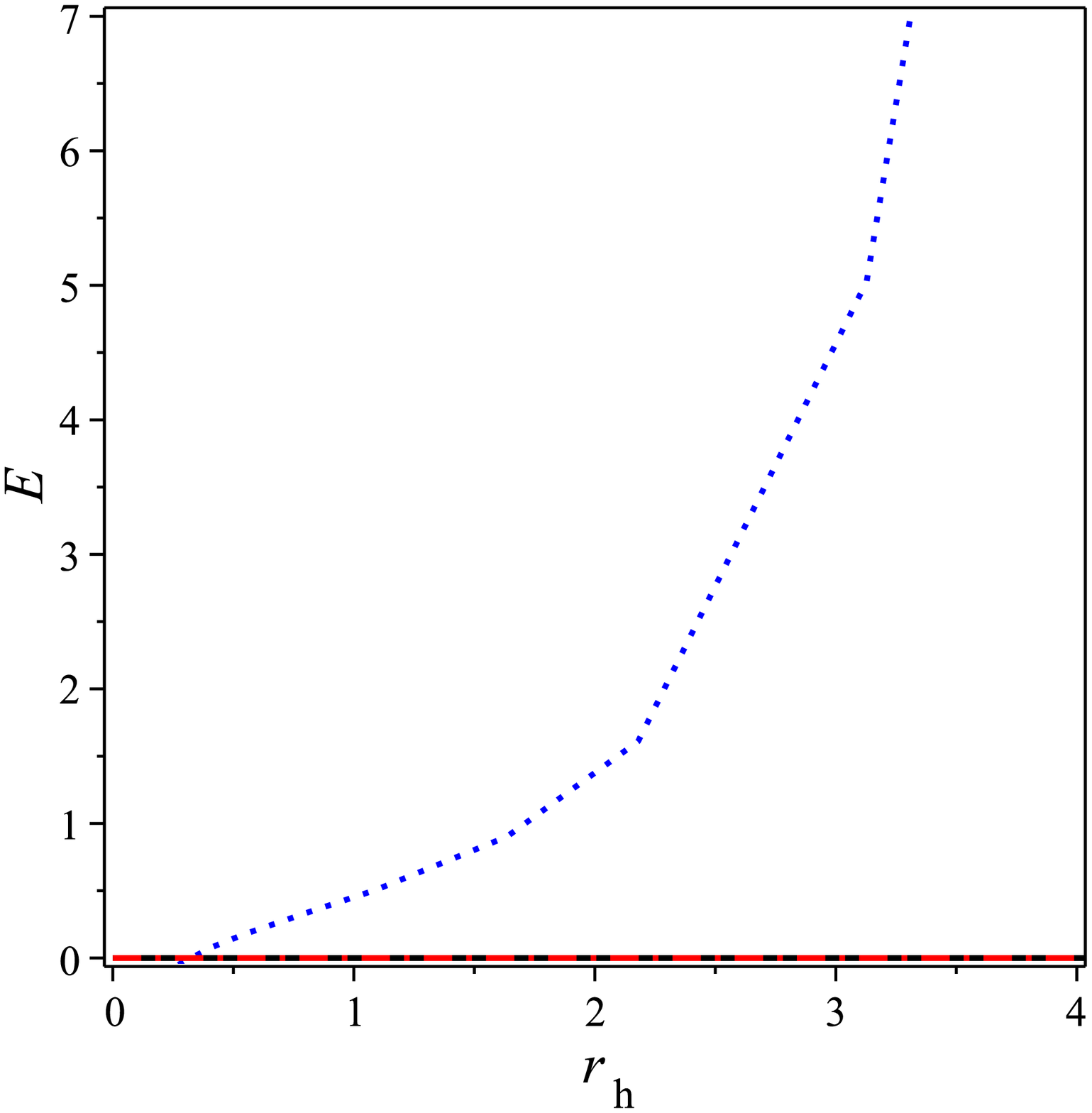}}
\subfigure[~Heat capacity of BH (\ref{mpab})]{\label{fig:hc}\includegraphics[scale=0.3]{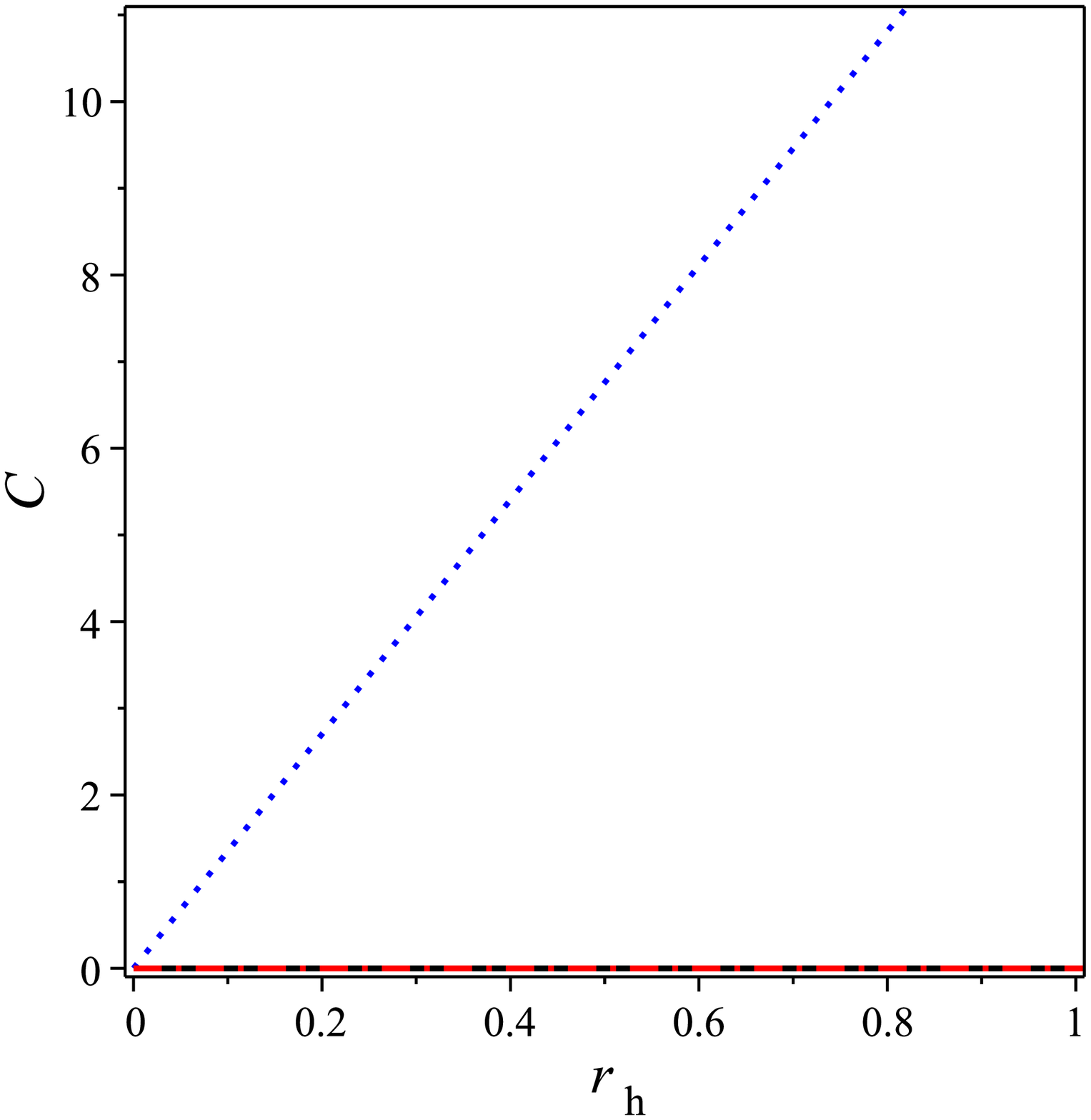}}
\subfigure[~Gibbs energy of BH (\ref{mpab})]{\label{fig:gib}\includegraphics[scale=0.3]{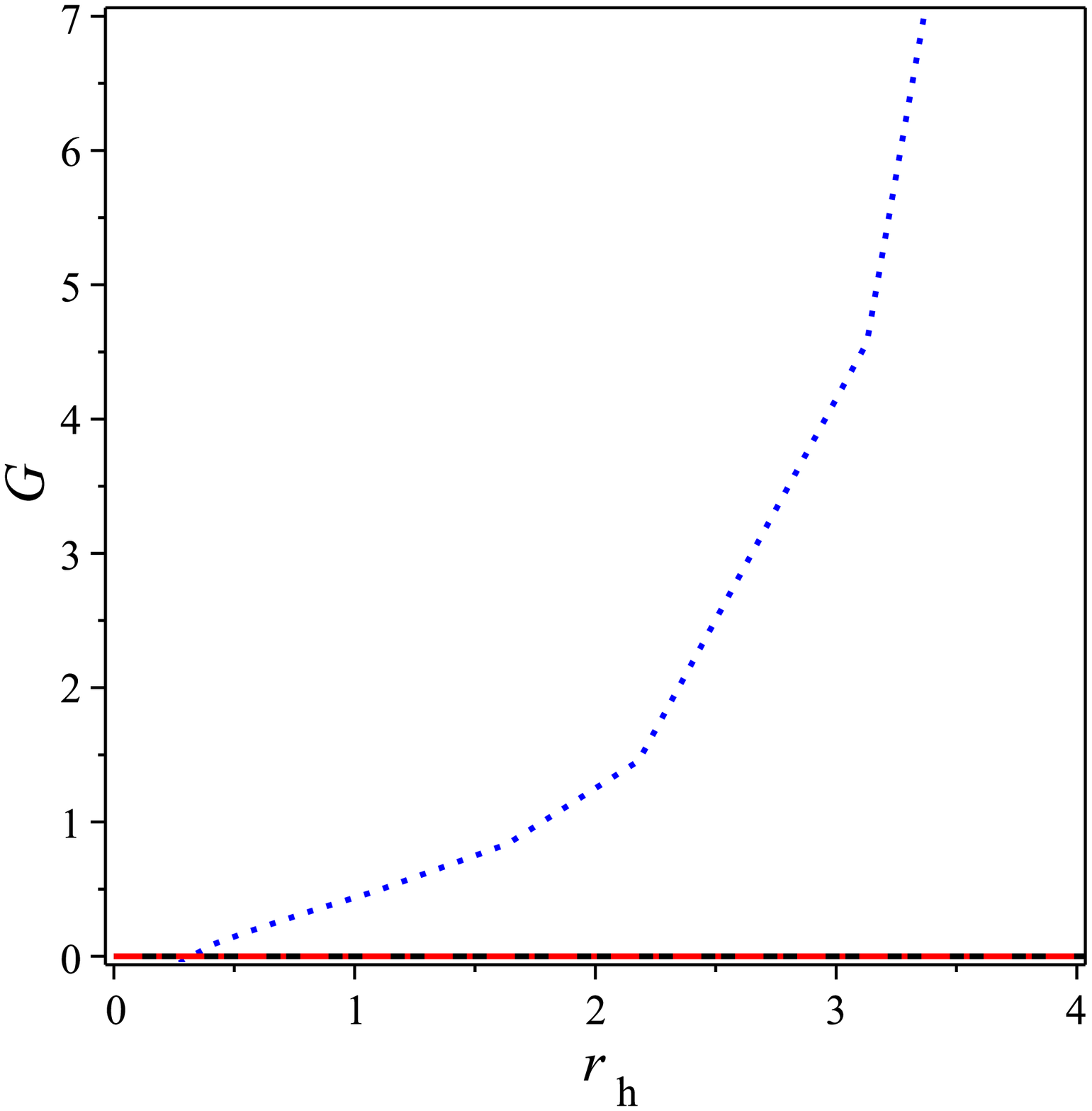}}
\caption[figtopcap]{\small{{Schematic plots of the thermodynamical quantities of the black hole solution  given by Eq. (\ref{metaf}): \subref{fig:ent} plot  of the entropy against  the radius horizon showing how the entropy consistently increases as the radius horizon increases; \subref{fig:Enr} typical behavior of the quasi--local energy given by (\ref{E1}); \subref{fig:hc} typical behavior of the horizon heat capacity, given by Eq. (\ref{hc}) and  \subref{fig:gib} typical behavior of the Gibb's free energy given by Eq. (\ref{Gib}). The quasi-local energy horizon heat capacity and Gibb's free energy all increase as $r_h$ increase. All of these values are plotted for $\Lambda_\mathrm{eff}=0.02$.}}}
\label{Fig:2}
\end{figure}

Using Eq. (\ref{ent}) we obtain the entropy of the BH  (\ref{mpab}) in the form
\begin{eqnarray}\label{S1}
S= \pi(a+r_h{}^2)\,.
\end{eqnarray}
According to Eq. (\ref{S1}), the entropy is different from the usual GR entropy due to the existence of the parameter $a$, when vanishes, we again obtain the usual  GR entropy. The difference results  from the  BH described here having a non-trivial value of the derivative of the function $f(R)$. The  behavior of the entropy is shown  in Fig. \ref{Fig:2} \subref{fig:ent} which indicates that $S$ increases as $r_h$ increases.

From Eq. (\ref{en}),  the quasi-local energy takes the form
\begin{eqnarray}\label{E1}
&&E=\frac{\Lambda_\mathrm{eff} [6r_h{}^6-5r_h{}^4+35ar_h{}^4+30a^2r_h{}^3-60ar_h{}^2-30a^2r_h\ln(r_h)+180a^2r_h{}^2+60a^3r_h\ln(r_h)-105a^3]-5a+5r_h{}^2}{10r_h}\,.\nonumber\\
&&
\end{eqnarray}
The behavior of the quasi--local--energy $E$ is shown in Fig. \ref{Fig:2} \subref{fig:Enr} which  shows that $E$ also increases as $R_h$ increases. Using Eq. (\ref{hc}) we obtain the heat capacity of the BH in the form
\begin{eqnarray}\label{hc1}
C=\frac{ 4\pi r_h{}^4(2r_h{}^2-5a)}{4r_h{}^4-8ar_h{}^2+15a^2}\,.
\end{eqnarray}
The behavior of the heat capacity is illustrated  in  Fig. \ref{Fig:2} \subref{fig:hc} which  shows that $C$ also increases as $r_h$ increases.

Finally, we substitute  Eqs. (\ref{T1}),  (\ref{S1}) and  (\ref{E1}) in Eq.  (\ref{enr}) to calculate the Gibbs free energy and obtain
\begin{eqnarray}\label{Gib}
&&G=\frac{1}{20r_h(2r_h{}^2-3a)}\Big(\Lambda_\mathrm{eff} r_h \Big[24r_h{}^7+104ar_h{}^5-50r_h{}^5-165ar_h{}^3+510a^2r_h{}^3+120a^2r_h{}^4-180a^3r_h{}^2+435a^2r_h\nonumber\\
&&-1500a^3r_h-30\ln(r_h)\{4a^2r_h{}^2-6a^3-8a^3r_h{}^2
+12a^4-21a^4\}
\Big]+10[3a^2+2r_h{}^4-5ar_h{}^2]\Big)\,.
\end{eqnarray}
 The behavior of this free energy is  illustrated  in Fig. \ref{Fig:2} \subref{fig:gib}; $G$  also increases as $r_h$ increases.

It was explained that the use of the thermalon procedure played an important role in the phase transition from AdS to dS \cite{Samart:2020qya}. Moreover, it is shown that  there are big correspondence  between Schwarzschild AdS/dS and  BH solutions corresponding dual CFTs living on the branes \cite{Nojiri:2002qn}.  Can the procedures applied in \cite{Samart:2020qya, Nojiri:2002qn} be done on the BHs derived in this study? At present we have no concrete answer. This needs more study which can be done in the future.
%%%%%%%%%%%%%%%%%%%%%%%%%%%%%%%%%%%% Section 3 %%%%%%%%%%%%%%%%%%%%%%%%%%%%%%%%%%%%%%%%
\subsection{Application of the first law of thermodynamics to the BH\lowercase{s}  described by Eq. (\ref{metaf}) }
%%%%%%%%%%%%%%%%%%%%%%%%%%%%%%%%%%%%%%%%%%%%%%%%%%%%%%%%%%%%%%%%%%%%%%%%%%%%%%%%%%%%%%
It is important to check that first law of thermodynamics is valid for the  BHs  given by Eq. (\ref{mpab}). Thus, we  apply this law to $f(R)$ using the  form \cite{Zheng_2018}
\begin{equation}
dE=TdS-PdV,\label{1st}
\end{equation}
where $E$ is the quasi-local energy, $S$ is the Bekenstein--Hawking entropy, $T$ is the hawking temperature, $P$ is the radial component of the stress-energy tensor that serves as the thermodynamic pressure $P=T_r{}^r\mid_{\pm}$ and $V$ is the geometric volume. Within the framework of $\mathrm{f(R)}$ gravitational theory, the pressure is defined as \cite{Zheng_2018}
\begin{equation}
P=-\frac{1}{8\pi}\left\{\frac{F}{r_h{}^2}+\frac{1}{2}(f-RF)\right\}+\frac{1}{4}\left(\frac{2F}{r_h}+F'\right)T\,.\label{11st}
\end{equation}
For the  spacetime described by (\ref{mpab}) if we neglect $O\Big(\frac{1}{r^4}\Big)$ to make the calculations more applicable we get\footnote{When we neglect the terms  of order $O\Big(\frac{1}{r^4}\Big)$ and higher orders and when $A(r)=0$ we obtain three roots, one of which is positive and two of which are imaginary.}

 \begin{eqnarray}\label{pr}
 P=-\frac{a+21a^3\Lambda_\mathrm{eff} -6a^2r_h\Lambda_\mathrm{eff}+12r_ha^3\Lambda_\mathrm{eff}-12ar_h{}^\Lambda_\mathrm{eff}+r_h{}^2+36a^2r_h{}^2\Lambda_\mathrm{eff}+
 12a^2r_h{}^3\Lambda_\mathrm{eff}}{2\,r_h{}^2}\,.
\end{eqnarray}
 The derivatives of   Eqs. (\ref{S1}) and (\ref{E1}) are
 \begin{eqnarray} \label{1stv}
&& dE=\frac{r_h{}^2+a+\Lambda_\mathrm{eff}[21a^3-3r_h{}^4-12ar_h{}^2-6a^2r_h+6r_h{}^6
+21ar_h{}^4+36a^2r_h{}^2+12a^2r_h{}^3
+12a^3r_h]}{2\,r_h{}^2}\,,\quad dS=2\pi r_h\,.\nonumber\\
&&
\end{eqnarray}
If we substitute  Eqs. (\ref{T1}), (\ref{pr}) and (\ref{1stv}) in (\ref{1st}) we can show that the first law of thermodynamics aapplies to the BHs.
%%%%%%%%%%%%%%%%%%%%%%%%%%%% Section 7 %%%%%%%%%%%%%%%%%%%%%%%%%%%%%
\section{Derivation of the stability of  the  BH\lowercase{s}  using the geodesic deviation}\label{S5}
%%%%%%%%%%%%%%%%%%%%%%%%%%%%%%%%%%%%%%%%%%%%%%%%%%%%%%%%%%%%%%%%%%%%

%%%%%%%%%%%%%%%%%%%%%%%%%%%% Section 7 %%%%%%%%%%%%%%%%%%%%%%%%%%%%%
%\subsection{Geodesic deviation}\label{S66}
%%%%%%%%%%%%%%%%%%%%%%%%%%%%%%%%%%%%%%%%%%%%%%%%%%%%%%%%%%%%%%%%%%%%
  The path of a  test particle in a gravitational field is described  by \cite{Nashed:2003ee},
 \begin{equation}\label{ge}
\mathit{ d^\mathrm{2} x^\sigma \over d\zeta^\mathrm{2}}+ \left\{^\sigma_{ \mu \nu} \right \}
 {d x^\mu \over d\zeta}{d x^\nu \over d\zeta}=0,
 \end{equation}
 where $\zeta$ is an affine  parameter along the geodesic. Equation (\ref{ge}) is the geodesic equation which can be derived as follows  \cite{1992ier..book.....D},
  \begin{equation} \label{ged}
 \mathit{{d^\mathrm{2} \xi^\sigma \over d\zeta^\mathrm{2}}+ \mathrm{2}\left\{^\sigma_{ \mu \nu} \right \}
 {d x^\mu \over d\zeta}{d \xi^\nu \over ds}+
 \left\{^\sigma_{ \mu \nu} \right \}_{,\ \rho}
 {d x^\mu \over d\zeta}{d x^\nu \over d\zeta}\xi^\rho=0},
 \end{equation}
where $\xi^\rho$ is the deviation 4-vector. Substituting Eqs. (\ref{ge}) and (\ref{ged})  into Eq.~(\ref{met12}) gives \begin{equation} \label{gedi}
\mathit{{d^\mathrm{2} t \over d\zeta^\mathrm{2}}=0, \qquad {1 \over \mathrm{2}} V'(r)\left({d t \over
d\zeta}\right)^\mathrm{2}-r\left({d \phi \over d\zeta}\right)^\mathrm{2}=0, \qquad {d^\mathrm{2}
\theta \over d\zeta^\mathrm{2}}=0,\qquad {d^\mathrm{2} \phi \over d\zeta^\mathrm{2}}=0,}\end{equation} and for
Eq. (\ref{ged}) we get \begin{eqnarray}\label{ged1} && \mathit{ {d^\mathrm{2} \xi^\mathrm{1} \over d\zeta^\mathrm{2}}+WV'{dt \over d\zeta}{d
\xi^\mathrm{0} \over d\zeta}-\mathrm{2}r W {d \phi \over d\zeta}{d \xi^\mathrm{3} \over
d\zeta}+\left[{\mathrm{1} \over \mathrm{2}}\left(V'W'+W(r) V''
\right)\left({dt \over d\zeta}\right)^\mathrm{2}-\left(W+rW'
\right) \left({d\phi \over d\zeta}\right)^\mathrm{2} \right]\xi^\mathrm{1}=0}, \nonumber\\
&& \mathit{ {d^\mathrm{2} \xi^\mathrm{0} \over
d\zeta^\mathrm{2}}+{V'\over V}{dt \over d\zeta}{d \xi^\mathrm{1} \over d\zeta}=0,\qquad {d^\mathrm{2} \xi^\mathrm{2} \over d\zeta^\mathrm{2}}+\left({d\phi \over d\zeta}\right)^\mathrm{2}
\xi^\mathrm{2}=0, \qquad \qquad  {d^\mathrm{2} \xi^\mathrm{3} \over d\zeta^\mathrm{2}}+{\mathrm{2} \over r}{d\phi \over d\zeta} {d
\xi^\mathrm{1} \over d\zeta}=0,} \end{eqnarray} where $V$ and $W$ are defined by   Eq.  (\ref{mpab}). Equations (\ref{gedi})  and (\ref{ged1}) are the geodesic and geodesic deviations of the line--element (\ref{met12}). Using
the circular orbit  \begin{equation} \mathit{\theta={\pi \over \mathrm{2}}, \qquad
{d\theta \over d\zeta}=0, \qquad {d r \over d\zeta}=0,}
\end{equation}
we get
\begin{equation}
 \mathit{\left({d\phi \over d\zeta}\right)^\mathrm{2}={V'
\over r(\mathrm{2}V-rV')}, \qquad \left({dt \over
d\zeta}\right)^\mathrm{2}={\mathrm{2} \over \mathrm{2}V-rV'}.} \end{equation}

Equation (\ref{ged1}) can also be written as
\begin{eqnarray} \label{ged2} &&  \mathit{{d^\mathrm{2} \xi^\mathrm{1} \over d\phi^\mathrm{2}}+WV' {dt \over
d\phi}{d \xi^\mathrm{0} \over d\phi}-\mathrm{2}r W {d \xi^\mathrm{3} \over
d\phi} +\left[{\mathrm{1} \over \mathrm{2}}\left(V'W'+W V''
\right)\left({dt \over d\phi}\right)^\mathrm{2}-\left(W+rW'
\right)  \right]\zeta^\mathrm{1}=0,} \nonumber\\
&&\mathit{{d^\mathrm{2} \xi^\mathrm{2} \over d\phi^\mathrm{2}}+\xi^\mathrm{2}=\mathrm{0}, \qquad {d^\mathrm{2} \xi^\mathrm{0} \over d\phi^\mathrm{2}}+{V' \over
V}{dt \over d\phi}{d \xi^\mathrm{1} \over d\phi}=\mathrm{0},\qquad {d^\mathrm{2} \xi^\mathrm{3} \over d\phi^\mathrm{2}}+{\mathrm{2} \over r} {d \xi^\mathrm{1} \over
d\phi}=\mathrm{0}.} \end{eqnarray}
The second equation in Eq.~(\ref{ged2}) describes a stable simple harmonic motion. Assuming the solutions to the rest of Eq. (\ref{ged2}) to have the form  \begin{equation} \label{ged3}
\mathit{\xi^\mathrm{0} = \varsigma_\mathrm{1} e^{i \omega \phi}, \qquad \xi^\mathrm{1}= \varsigma_\mathrm{2}e^{i \omega
\phi}, \qquad and \qquad \varsigma^\mathrm{3} = \zeta_\mathrm{3} e^{i \omega \phi},}\end{equation} where
$\varsigma_1$, $\varsigma_2$ and $\varsigma_3$ are constants and    $\phi$   must be determined. From Eqs.  (\ref{ged3}) and
(\ref{ged2}) we have  \begin{equation} \label{con1}  \mathit{\displaystyle\frac{\mathrm{3}VWV'-\omega^\mathrm{2}VV'-\mathrm{2}rWV^\mathrm{'2}+rVWV''}{VV'}>0,} \end{equation} which is the stability condition. The solution to Eq.~(\ref{con1}) has the form  \begin{equation}\label{stc3}\omega^2=\mathit{\displaystyle\frac{\mathrm{3}VWV^\mathrm{'}-\mathrm{2}rWV^\mathrm{'2}+rVWV''}{VV'}>0,}\end{equation}
which is the  stability condition for the Eq. (\ref{mpab}) \cite{Misner:1974qy}. Substituting the metric potentials given by Eq.~(\ref{mpab}) in Eq.~(\ref{stc3}) we can obtain the behavior of the stability condition which is also shown in Fig. 3.
\begin{figure}
\centering
\includegraphics[scale=0.3]{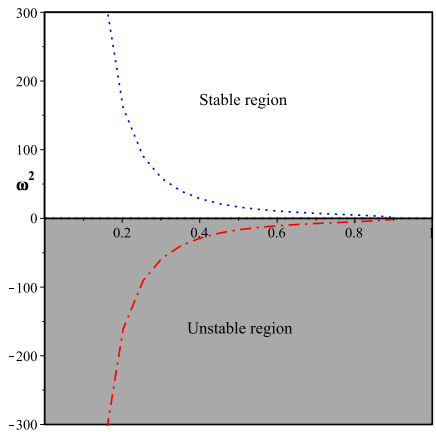}
\caption[figtopcap]{\small{{Schematic plot of the stability condition described by Eq. (\ref{stc3})showing the stable and unstable regions.}}}
\label{Fig:3}
\end{figure}

\section{Summary and conclusions }\label{S77}
Generally, the field equations of $\mathrm{f(R)}$ are complicated  and it is not easy to find an analytical solution. By using the trace of the field equations we can write $\mathrm{f(R)}$ in terms of the remaining terms, i.e., in terms of the derivatives of $\mathrm{f(R)}$ and the  d'Alembertian operator. Using this form of $\mathrm{f(R)}$  we can  rewrite  the field equations and apply them to flat horizons using two unknown functions of the radial coordinate. We then derived the differential equations and analyzed them for special cases  and derived special cases that coincided with the GR BHs. From this analysis we then derived  general non-trivial BHs that are different from the BHs of GR. These BHs are characterized by a convolution function and depend on a constant that is responsible for making these BHs deviate from GR BHs. In spite of the fact that  the field equations  do not include a cosmological constant, we obtained solutions that included an effective cosmological constant, this is an advantage  of $\mathrm{f(R)}$  gravitational theory. By using  coordinate transformations between the temporal and $\phi$ coordinates we succeeded in deriving new forms of rotating BHs with non-trivial values of the Ricci scalar.

To understand the physics of these BHs we derived the forms of their asymptotes form and showed that they behaved as AdS/dS depending on the sign of the effective cosmological constant. We also showed that such BH solutions coincide with GR BHs when the constant that is associated with the derivative of  $\mathrm{f(R)}$  equals zero. Also, we derived the asymptotic form  of $\mathrm{f(R)}$  for these solutions  and showed that it behaves as a polynomial function. To check the singularities of these solutions, we calculated the invariants and showed that the higher--order--curvature terms make the singularities of these BHs  softer than those of GR BHs.  Thermodynamical quantities such as horizons, Hawking temperature, entropy, heat capacity and Gibbs free energy were also calculated and it was shown that their behavior is consistent with that described in the literature. Moreover, it was demonstrated  that these solutions satisfy the first law of thermodynamics. To test the stability of the  BHs we used the geodesic deviation and derived the stability condition  analytically, the regions of stability were also illustrated graphically. Finally,  using  odd--type perturbations methods  \cite{Nashed:2019tuk,Elizalde:2020icc} we  showed that our BHs have no ghosts and that the radial speed has a value of one, which insures that these BHs are stable.

In conclusion we stress that our derived BHs are different from those of GR due to the constant  coefficient  in the derivative of  $\mathrm{f(R)}$ which is of order two. Indeed if we change the order of $r$  such that is its coefficient is a constant we derive new BHs that may correspond  to a physics completely different  from the BHs presented in this study. This is a topic future study.
%\begin{acknowledgments}
%This work is partially supported  by MEXT KAKENHI Grant-in-Aid for Scientific Research on Innovative Areas ``Cosmic Acceleration'' No. 15H05890 (S.N.) and the JSPS Grant-in-Aid for Scientific Research (C) No. 18K03615 (S.N.).
%\end{acknowledgments}
%%%%%%%%%%%%%%%%%%%%%%%%%%%%%%%%%
%\bibliographystyle{apsrev}
%\bibliography{JRPHSRef}
%%%%%%%%%%%%%%%%%%%%%%%%%%%%%%%%%%%%%%%%%%%%%%%%%%%%%%%%%%%%%%%%%%%%%%%%%%%%%%%%%%%%%%
%merlin.mbs apsrev4-1.bst 2010-07-25 4.21a (PWD, AO, DPC) hacked
%Control: key (0)
%Control: author (8) initials jnrlst
%Control: editor formatted (1) identically to author
%Control: production of article title (-1) disabled
%Control: page (0) single
%Control: year (1) truncated
%Control: production of eprint (0) enabled
%

\end{document}